\documentclass[11pt,reqno]{amsart}

\usepackage[T1]{fontenc}     
\usepackage{lmodern}         
\usepackage{amsmath, amsthm, amssymb,amsaddr}
\usepackage{dcolumn}
\usepackage{bm}
\usepackage{amsmath} 
\usepackage{mathrsfs}
\usepackage{mathtools}
\usepackage{algpseudocode}
\usepackage{algorithm}
\usepackage[pdftex]{graphicx}
\usepackage{subfigure}
\usepackage{epstopdf}
\usepackage{tikz}
\usepackage{xcolor}
\usepackage[export]{adjustbox}
\usepackage{caption}
\usepackage{hyperref}
\usepackage[pagewise]{lineno}
\usepackage{multirow}
\usepackage{caption}
\usepackage{tikz}
\usepackage{subcaption}
\usepackage[export]{adjustbox} 

\theoremstyle{remark}

\newif\ifblacktext
\blacktextfalse 

\ifblacktext
\fi

\newcommand{\be}[1]{\begin{equation}\label{#1}}
\newcommand{\ee}{\end{equation}}

\newcommand{\no}[1]{#1}

\newcommand{\bx}{\boldsymbol{x}}

\newcommand{\iu}{{i\mkern1mu}}

\renewcommand{\no}[1]{} 

\title[Hessian-Free Ray-Born Inversion for QUT: In Vitro and In Vivo Validation]{The First In Vitro and In Vivo Validation of the Hessian-Free Ray-Born Inversion for Quantitative Ultrasound Tomography}

\author{Ashkan Javaherian}
\address{Department of Bio-Electric, School of Electrical and Computer Engineering, University College of Engineering, University of Tehran, Tehran, Iran.
\footnote{The \textsc{MATLAB} codes supporting the findings of this study, as well as those in \cite{Javaherian_2022,Javaherian_2021,Javaherian_2020}, are publicly available at the following GitHub repository:
\url{https://github.com/Ash1362/ray-based-quantitative-ultrasound-tomography/}.
The in vitro and in vivo datasets used in this work are available at:
\url{https://github.com/rehmanali1994/WaveformInversionUST/}.}}
\email{ajavherian62@gmail.com; ashkan.javaherian@ut.ac.ir}
\date{Novemeber 2025}

\begin{document} 
\maketitle

\begin{abstract}
This study presents the first experimental validation of a Hessian-free ray-Born inversion technique for quantitative reconstruction of sound speed from transmission ultrasound data. The method combines single-scattering theory with high-frequency approximations, yielding an inversion framework well suited to the frequency ranges used in clinical ultrasound applications. Unlike previous singly-scattered inversion approaches that account for medium heterogeneities only in the scattering potential, the proposed ray-Born method employs Green’s functions approximated along ray trajectories determined by high-frequency assumptions. The associated objective function is linearized and minimized sequentially across increasing frequency bands. At each frequency set, the linearized subproblem is solved using a weighting scheme applied to both the solution and data spaces, which diagonalizes the Hessian and enables its inversion in a single step. The method, previously reported and released as an open-source package, was applied to \textit{in-vitro} and \textit{in-vivo} datasets provided by the University of Rochester Medical Center. The reconstructed images were evaluated by comparison with those obtained using a full-wave inversion approach based on a frequency-domain Helmholtz solver. The results demonstrate the strong potential of the Hessian-free ray-Born inversion as a computationally efficient and accurate method suitable for clinical translation.
\end{abstract}


\section{INTRODUCTION}
\label{sec:intro}  

Ray-based image reconstruction algorithms, which rely on singly scattered waves for image formation, have recently been proposed for quantitative ultrasound tomography (QUT), specifically for the reconstruction of sound-speed maps from transmission ultrasound data \cite{Javaherian_2022,Javaherian_2021}. Numerical studies have demonstrated that for weakly heterogeneous media, such as the breast tissue, these approaches can reconstruct high-resolution sound speed images from transmission ultrasound  data at a lower computational cost compared to commonly used full-waveform inversion techniques \cite{Wang,Wiskin,FAN202210,Robins,Ali}.  

In this work, we present the first validation results for the ray-Born inversion technique previously proposed in \cite{Javaherian_2022}. The reconstruction proceeds by sequentially linearizing and minimizing the objective function from low to high frequencies. At each frequency set, the objective function is linearized, and the resulting reconstructed image serves as the initial guess for the subsequent higher-frequency set \cite{Javaherian_2022,Javaherian_2021}. Solving the linear subproblem at each frequency effectively computes the action of the inverse Hessian on the gradient of the objective function \cite{Javaherian_2022,Javaherian_2021}. A numerical implementation and validation of the proposed ray-Born approaches for QUT of soft tissues using synthetic data have been made available as an open-source MATLAB package \cite{Javaherian_toolbox}.

Typically, ray-tracing approaches for quantitative sound speed imaging account for phase aberrations using a simplified or approximate model that includes refraction effects along rays connecting emitter–receiver pairs. In the forward model, the time of flight for each emitter–receiver pair is approximated by assuming that the acoustic waves propagate solely along the direct connecting ray \cite{Javaherian_2020,LI200961,GEMMEKE201759,QU20222079,Ceccato10923741}. In contrast, the ray-based forward model employed in our proposed inversion methods \cite{Javaherian_2022,Javaherian_2021} goes beyond refraction-induced phase aberrations, incorporating all parameters of the Green’s function, including geometric spreading, acoustic absorption, and dispersion \cite{Javaherian_2021}. Crucially, it also integrates singly-scattered waves into the reconstruction process by embedding ray tracing under a high-frequency assumption within the Born inversion framework \cite{Javaherian_2022,Javaherian_2021,Thierry}.

From an inverse problem perspective, the original objective function—which represents the \(L_2\)
norm of the discrepancy between a measured dataset and one approximated using a ray-based forward model \cite{Javaherian_2021}—is nonlinear. While the measured dataset is fixed and obtained experimentally, the forward model is iteratively updated until the objective function is sufficiently minimized. In this model, rays are initialized at the emitters, propagate through the object immersed in water, and are intercepted by receivers, accounting for accumulated parameters along the rays while neglecting scattering effects \cite{Javaherian_2021}. For each frequency set, a Taylor-series expansion of the objective function, neglecting second- and higher-order terms, yields a linear equation \cite{Javaherian_2021}. Minimizing this linearized objective function provides a search direction equivalent to the implicit action of the inverse Hessian on the gradient of the original objective function \cite{Javaherian_2021}. This second-order formulation enables the incorporation of singly-scattering effects into the minimization process \cite{Javaherian_2021,Thierry}. The resulting minimization is equivalent to a form of the Born approximation, in which the forward model is described using Green’s functions whose parameters are no longer assumed homogeneous but are instead approximated along rays using a high-frequency approximation \cite{Javaherian_2021}. Progressing from low to high frequencies, the image reconstructed after solving the linear subproblem for one frequency set serves as the initial guess for the linear subproblem corresponding to the next frequency set.

While the inversion approach proposed in \cite{Javaherian_2021} solves the linear equation associated with each frequency set iteratively using a Conjugate Gradient (CG) algorithm, the method introduced in \cite{Javaherian_2022} applies a weighting scheme to both the data and solution spaces, allowing the Hessian matrix to be diagonalized and inverted in a single step. In this study, we adopt the latter inversion approach to reconstruct sound speed maps from in vitro and in vivo ultrasound datasets.

This work presents the first validation results obtained using the Hessian-free ray-Born inversion approach \cite{Javaherian_2022}. For this purpose, the open-source MATLAB package for ray-based transmission ultrasound tomography \cite{Javaherian_toolbox} was used to reconstruct quantitative sound speed images from transmission ultrasound tomography datasets recently released by the University of Rochester Medical Center (URMC) \cite{Ali,urmc_data}

Section 2 briefly outlines the ray-based forward model. Section 3 derives the associated objective function and describes the linearization and minimization process; further methodological details can be found in \cite{Javaherian_2022,Javaherian_2021}. Section 4 presents the images reconstructed from the in vivo and in vitro ultrasound datasets made publicly available by the University of Rochester Medical Center (URMC) \cite{urmc_data}. These reconstructions are compared with those obtained using a full-waveform inversion solver based on the frequency-domain Helmholtz equation \cite{Ali}. Section 5 provides a brief discussion of the results.

\section{Forward Problem} \label{sec:forward}

The acoustic pressure field $p(\omega, \bx)$ at a single frequency $\omega$, generated by a point source $s$ located at $\bx_e$, is modeled using the Helmholtz equation \cite{Javaherian_2022,Javaherian_2021}:
\begin{align} \label{eq:wave-equation}
    \big[\nabla^2 + k^2 \big] \, p(\omega, \bx, \bx_e) = -s(\omega) \, \delta(\bx - \bx_e),
\end{align}
where $k = \omega / c$ is the wavenumber and $c$ denotes the spatially varying sound speed.

For a point source with a frequency-dependent amplitude $s(\omega)$, the solution to Eq.~\eqref{eq:wave-equation} can be expressed using Green's formula:
\begin{align} \label{eq:greens-formula}
    p(\omega, \bx) = g(\omega, \bx, \bx_e) \, s(\omega, \bx_e),
\end{align}
where the integration over the source support is omitted by assuming the source is restricted to a point. The Green's function $g(\omega, \bx, \bx_e)$ satisfies the homogeneous Helmholtz equation:
\begin{align} \label{eq:wave-equation-hom}
    \big[\nabla^2 + k^2 \big] \, g(\omega, \bx, \bx_e) = -\delta(\bx - \bx_e).
\end{align}

In a 2D medium, the Green's function can be approximated as:
\begin{align} \label{eq:greens}
    g(\omega, \bx, \bx_0) \approx A(\bx, \bx_0) \exp \big(\iu \left[ \phi(\bx, \bx_0) + \pi/4 \right] \big),
\end{align}
where $\phi(\bx, \bx_0)$ denotes the accumulated phase, and $A(\bx, \bx_0)$ denotes the amplitude decay, expressed as
\begin{align}
    A(\bx, \bx_0) = A_{\rm geom} (\bx, \bx_0) \, A_{\rm abs} (\bx, \bx_0),
\end{align}
with $A_{\rm geom}$ and $A_{\rm abs}$ representing the amplitude decay due to geometrical spreading and accumulated acoustic absorption, respectively \cite{Javaherian_2022,Javaherian_2021}.

Substituting Eq.~\eqref{eq:greens} into the homogeneous wave equation \eqref{eq:wave-equation-hom} and applying the high-frequency approximation yields the \textit{Eikonal} and \textit{Transport} equations. The Eikonal equation, which governs ray trajectories and accumulated phase, is given by \cite{Javaherian_2022,Javaherian_2021}:
\begin{align} \label{eq:eikonal}
    \nabla \phi \cdot \nabla \phi = k^2.
\end{align}
The Transport equation, which determines amplitude decay due to geometrical spreading, is given by \cite{Javaherian_2022,Javaherian_2021}:
\begin{align} \label{eq:transport}
    \nabla \cdot \big(A^2 \nabla \phi \big) = 0.
\end{align}

Equations~\ref{eq:eikonal} and~\ref{eq:transport} together describe the propagation of acoustic waves in a medium, capturing both phase and amplitude aberrations along ray trajectories.

\section{Inverse Problem} \label{sec:inverse}

The inverse problem aims to estimate the spatially varying sound speed $c$ that minimizes the following objective function:
\begin{align} \label{eq:objective}
    F(c) = \frac{1}{2} \sum_{e,r} \int | p_{(\omega, r, e)}(c) - \hat{p}_{(\omega, r, e)} |^2 \, d\omega,
\end{align}
where $p_{(\omega, r, e)}(c)$ is the pressure field approximated using the ray-based forward model at receiver $r$ after excitation from emitter $e$ at frequency $\omega$. The term $\hat{p}_{(\omega, r, e)}$ represents the measured pressure data obtained experimentally.  

Using the forward operator described in Section~\ref{sec:forward}, the modeled pressure field $p_{(\omega, r, e)}(c)$ incorporates only the accumulated information along the rays linking emitter-receiver pairs, neglecting scattering effects \cite{Javaherian_2021}. Consequently, a direct minimization of the objective function \eqref{eq:objective} results in the reconstruction of a low-resolution sound speed map $c$.  

To overcome this limitation, the problem is reformulated to compute updates, $\delta c$, that minimize a Taylor series expansion of the objective function
\begin{align} \label{eq:objective-taylor}
    F(c + \delta c) = F(c) + \int \frac{\partial F}{\partial c(\bx')} \delta c(\bx') \, d\bx' + O\big((\delta c)^2\big).
\end{align}
The update step $\delta c$ that minimizes this expanded objective function is derived by computing and zeroing the gradient of the expanded objective function with respect to $c$. At iteration $n$, this results in a linearized equation for the descent step $\delta c^{(n)}$ \cite{Javaherian_2021}:
\begin{align} \label{eq:linear}
    \begin{split}
        \frac{\partial F(c^{(n)} + \delta c^{(n)})}{\partial c^{(n)}} &\approx \frac{\partial F(c^{(n)})}{\partial c^{(n)}(\bx)} + \int \frac{\partial^2 F(c^{(n)})}{\partial c^{(n)}(\bx) \partial c^{(n)}(\bx')} \delta c^{(n)}(\bx') \, d\bx', \\
        &\approx \nabla F^{(n)}(\bx) + \int H^{(n)}(\bx, \bx') \delta c^{(n)}(\bx') \, d\bx' = 0.
    \end{split}
\end{align}

In Eq.~\eqref{eq:linear}, $\nabla F^{(n)}$ and $H^{(n)}$ denote the first-order and second-order derivatives of $F$ with respect to $c^{(n)}$, also referred to as the \textit{gradient} and the \textit{Hessian matrix}, respectively.

By minimizing the Taylor-series expanded objective function \eqref{eq:objective-taylor}, the update directions $\delta c^{(n)}$ are determined from the linearized equation \eqref{eq:linear}, which incorporates scattering effects through the actions $H^{(n)} \delta c^{(n)}$. Starting with a smooth initial guess, typically obtained using early iteration of a time-of-flight-based reconstruction algorithm \cite{Javaherian_2020}, and iteratively solving Eq.~\eqref{eq:linear} across increasing frequencies, the scattering features are progressively incorporated. This iterative approach progressively refines the reconstructed image, enabling the accurate capture of small-scale features essential for high-resolution quantitative imaging.

The linear equations~\eqref{eq:linear} are typically solved iteratively using a Conjugate Gradient (CG) algorithm, requiring approximately 10--15 CG iterations for each frequency set. Each iteration involves computing the action of the Hessian on the sound speed map, \(H c\), which substantially increases the computational cost. Moreover, the convergence behavior of the CG algorithm is frequency-dependent and is affected by factors such as noise and initial guess.

Motivated by these limitations and time-domain studies using backscattered wavefields in the geophysics literature \cite{Thierry}, a single-step approach for solving the linear equations has been proposed \cite{Javaherian_2022}. Rather than solving each linear equation iteratively, this method applies weighting in both the data and solution spaces to diagonalize the Hessian matrix, enabling its inversion in a single step. Readers are referred to \cite{Javaherian_2022} for a detailed discussion of this image reconstruction approach and for an explanation of how the diagonalization introduces regularizing effects, resulting in more accurate reconstructions in the presence of noise or when initial guesses are far from the true solution.


\section{Results}

This section presents the sound speed images reconstructed using the ray-Born inversion approach proposed in \cite{Javaherian_2022}. The forward and inverse problems are briefly outlined in Sections~\ref{sec:forward} and~\ref{sec:inverse}, respectively, with further details provided in \cite{Javaherian_2022,Javaherian_2021}. The reconstructed images are compared with those obtained using a full-waveform inversion method based on the Helmholtz equation solved in the frequency domain \cite{Ali}. The ray-Born inversion approach described in \cite{Javaherian_2022} was employed to reconstruct quantitative sound speed images from the \textit{in vitro} and \textit{in vivo} datasets released as open source by the University of Rochester Medical Center (URMC) \cite{Ali,urmc_data}. Image reconstructions using the ray-Born approach were performed on a single 6-core Intel Xeon E5-2620 v2 2.1~GHz CPU. For benchmarking purposes, the images reconstructed using the full-waveform inversion method were extracted from the open-source dataset package \cite{urmc_data}.

\subsection{\textit{In Vitro} Phantom Datasets}

The measurement setup consisted of an array of 256 transducers arranged in a circular ring with a radius of 11~cm. Each transducer was sequentially excited, emitting acoustic waves that were recorded over time by the remaining transducers acting as receivers. This procedure was repeated for all transducers to complete a full dataset. The data acquisition protocol follows Section~III.B of \cite{Ali}.

The datasets used in this study are part of the open-source transmission ultrasound tomography dataset package \cite{urmc_data}. The sound speed in water, measured during the experiment, was 1480~m/s. Ultrasound data were acquired for two slices of a tissue-mimicking phantom, as described in Section~IV.C of \cite{Ali}.

The image reconstruction procedure using the full-wave approach is described in \cite{Ali}, whereas the proposed ray-Born inversion approach is detailed in \cite{Javaherian_2022}. The image reconstructed using the full-wave inversion approach, which serves as the benchmark, was computed on a grid with a spacing of 0.5~mm \cite{Ali}. In contrast, the image reconstructed using the ray-Born inversion approach proposed in \cite{Javaherian_2022} was computed on a grid with a spacing of 1~mm, demonstrating its computational scalability. 

For each emitter, only the receivers located on the opposite 75\%  of the ring were included in the image reconstruction to mitigate directivity effects arising from closely spaced emitter--receiver pairs. In addition, following \cite{Ali}, a single-sided Gaussian window was applied to each time trace to suppress events occurring before the first-arrival time.

While the full-wave inversion approach utilized frequencies ranging from 350~kHz to 650~kHz, the ray-Born reconstruction was performed over a wider range of 350~kHz to 950~kHz, both progressing incrementally from lower to higher frequencies. In the ray-Born reconstruction, each linear equation~\eqref{eq:linear} was formulated and solved at two consecutive frequencies. Consequently, the reconstruction comprised 70 linearization steps, corresponding to 140 frequencies within the supported range. For each frequency set, the largest 1\% of values were set to zero, as described in Section III.E of \cite{Ali}.

While the full-waveform inversion approach started from a homogeneous initial guess, the ray-based approach employed a low-resolution, low-contrast sound speed image reconstructed from the early iterations (here, three) of a time-of-flight (ToF)-based reconstruction method. This initial estimate was obtained by iteratively minimizing the discrepancy between the measured ToFs—extracted via a ToF-picking procedure—and the ToFs approximated along the rays \cite{Javaherian_2020,LI200961}.

\subsubsection{Slice (a) of the phantom}
\begin{figure} [ht]
\begin{center}
    \subfigure[]{\includegraphics[width= 0.33 \textwidth]{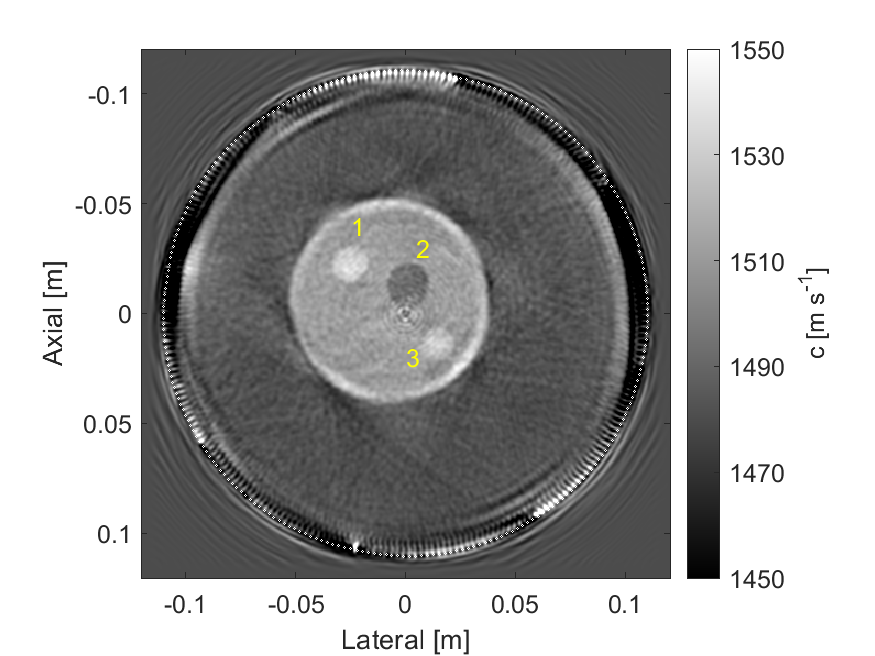} \label{fig:1a}}
    \subfigure[]{\includegraphics[width= 0.33 \textwidth]{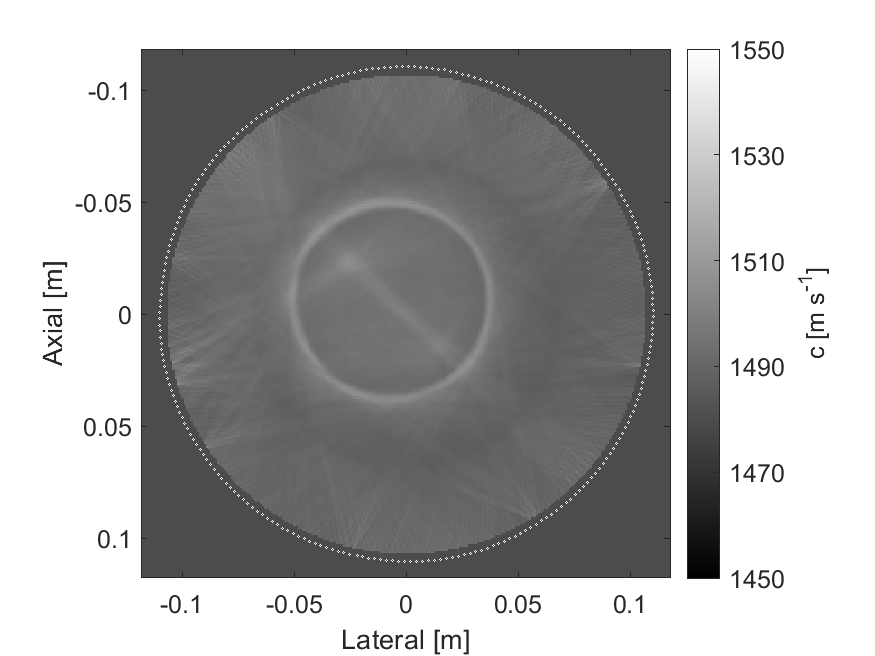}\label{fig:1b}}
    \subfigure[]{\includegraphics[width= 0.33 \textwidth]{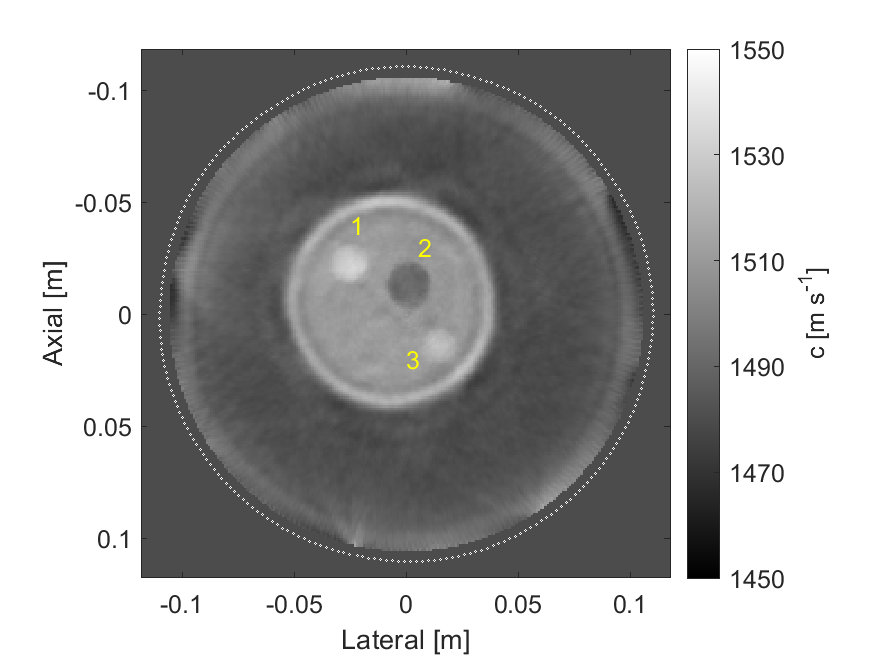} \label{fig:1c}}
    \subfigure[]{\includegraphics[width= 0.33 \textwidth]{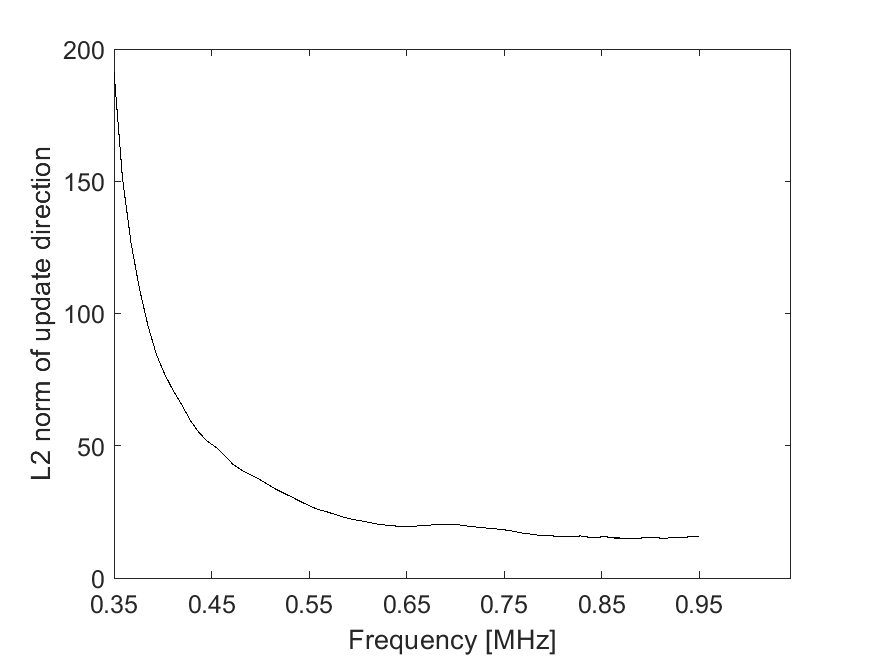} \label{fig:1d}}
   \subfigure[]{\includegraphics[width= 0.33 \textwidth]{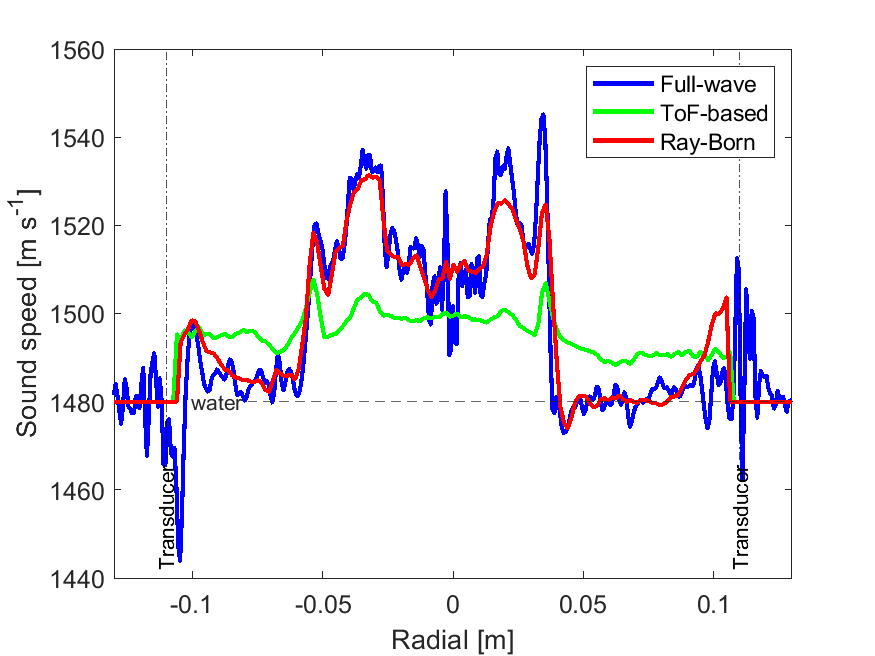} \label{fig:1e}}  
      \subfigure[]{\includegraphics[width= 0.33 \textwidth] {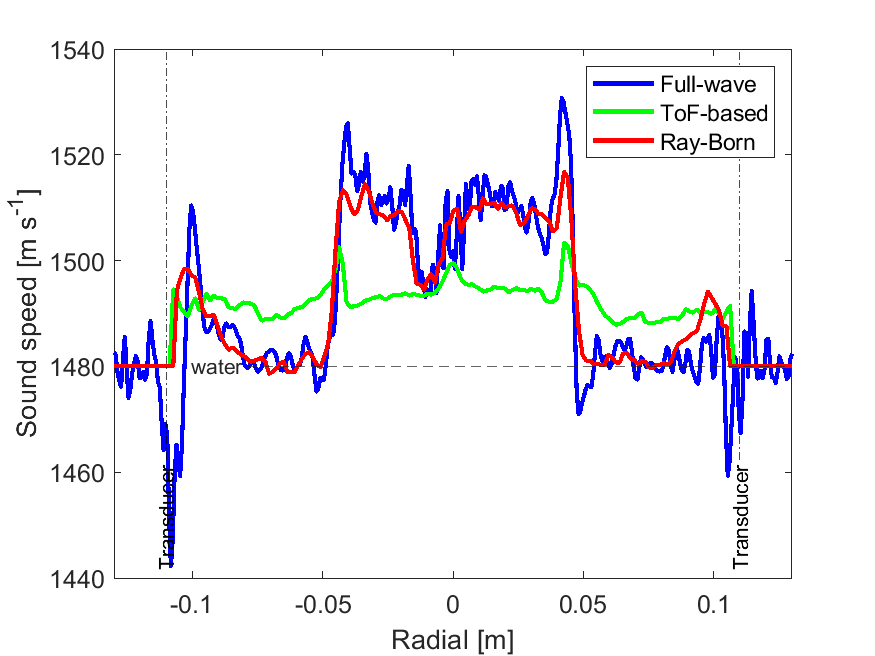} \label{fig:1f}}
       \subfigure[]{\includegraphics[width= 0.33 \textwidth]{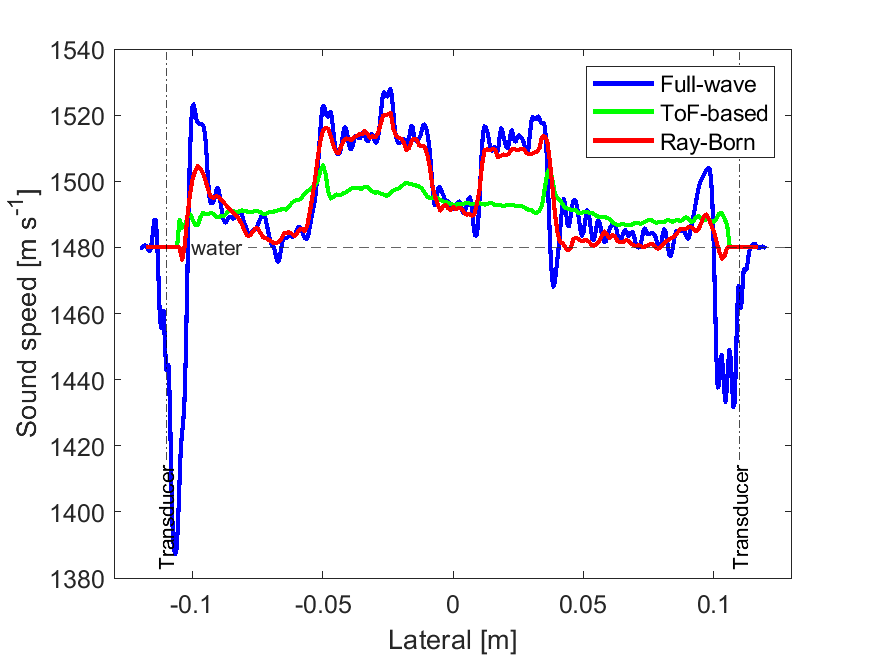} \label{fig:1g}}  
          \subfigure[]{\includegraphics[width= 0.33 \textwidth]{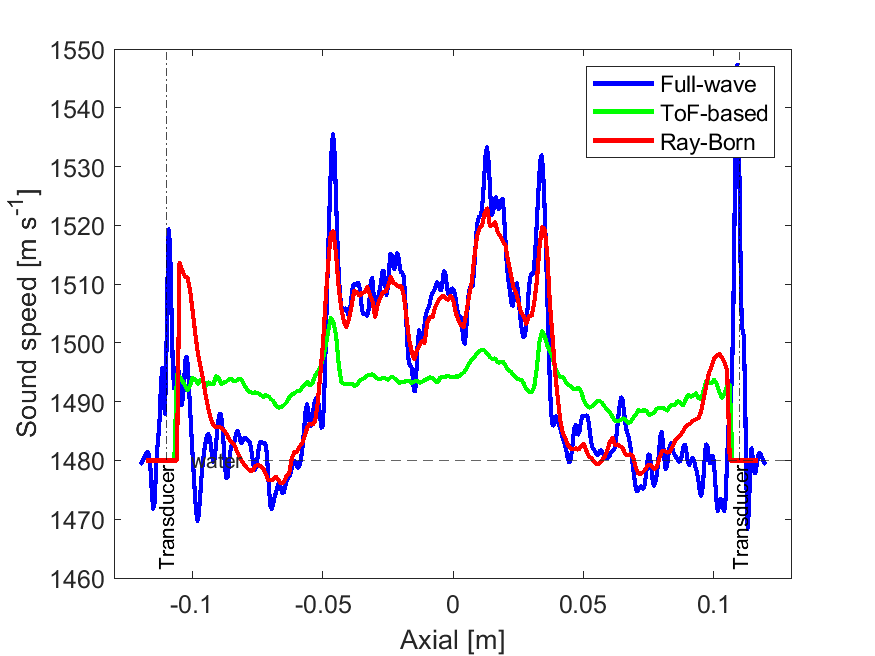} \label{fig:1h}}  
\end{center}
\caption{ \label{fig:1}
Sound speed images from the \textit{in vitro} dataset corresponding to slice~(a) of the phantom, reconstructed using: (a) full-wave inversion, (b) the time-of-flight (ToF)-based approach, which served as the initial guess, and (c) the ray-Born inversion approach. (d) The \(L_2\)-norm of the update direction in terms of sound speed for each frequency subproblem of the implemented ray-Born inversion. Line profiles of the reconstructed sound speed images along: (e) the first diagonal, (f) the second diagonal, (g) the \(x = -1.5 \ \mathrm{cm}\) (lateral) axis, and (h) the \(y = 1 \ \mathrm{cm}\) (axial) axis.
}
\end{figure}

Figure~\ref{fig:1} presents the reconstructed sound speed images from slice~(a) of the phantom described in Section~IV.C of \cite{Ali}. This figure illustrates the quantitative results obtained using the ray-Born inversion and full-waveform inversion approaches, allowing a direct visual comparison of their performance on the same experimental dataset.

Figure~\ref{fig:1a} illustrates the sound speed image reconstructed using the full-waveform inversion approach, as described in \cite{Ali,urmc_data}. Figure~\ref{fig:1b} shows the low-resolution, low-contrast image reconstructed using three linearizations of the ToF-based inversion method, which served as the initial guess for the ray-Born inversion. Figure~\ref{fig:1c} presents the sound speed image reconstructed using the ray-Born inversion technique proposed in \cite{Javaherian_2022}. A visual comparison between Figures~\ref{fig:1a} and~\ref{fig:1c} demonstrates that the ray-Born inversion approach successfully reconstructs all three inclusions (numbered~1--3) with quantitatively consistent sound speed values.

Figure~\ref{fig:1d} depicts the \(L_2\) norm of the update directions, i.e., the solutions \(\delta c\) of the 70 linear equations associated with 140 frequencies within the selected range, demonstrating the stable convergence of the ray-Born inversion algorithm and its robustness with respect to the choice of termination frequency.

Figures~\ref{fig:1e}--\ref{fig:1h} show the line profiles of the reconstructed sound speed images along the first diagonal, second diagonal, \(x=-1.5 \ \mathrm{cm}\) (lateral) axis, and \(y=1 \ \mathrm{cm}\) (axial) axis, respectively. The blue, green, and red curves correspond to the images reconstructed using the full-waveform, ToF-based, and ray-Born inversion approaches, respectively. The horizontal dotted line denotes the sound speed in water. It should be noted that the full-waveform reconstruction is not a ground truth but is included solely for benchmarking purposes.

\subsubsection{Slice (b) of the Phantom}

\begin{figure} [ht]
\begin{center}
    \subfigure[]{\includegraphics[width= 0.33 \textwidth]{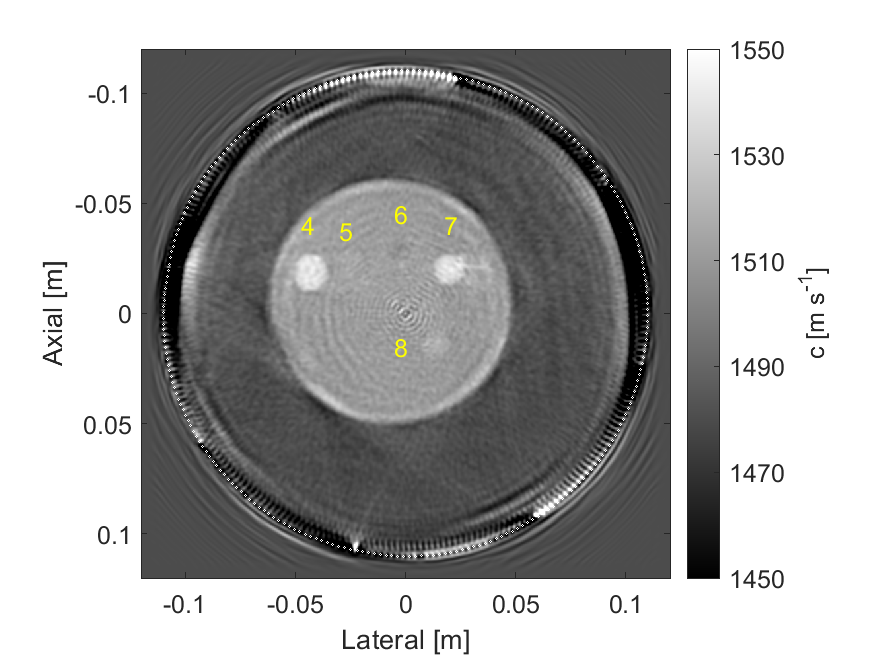} \label{fig:2a}}
    \subfigure[]{\includegraphics[width= 0.33 \textwidth]{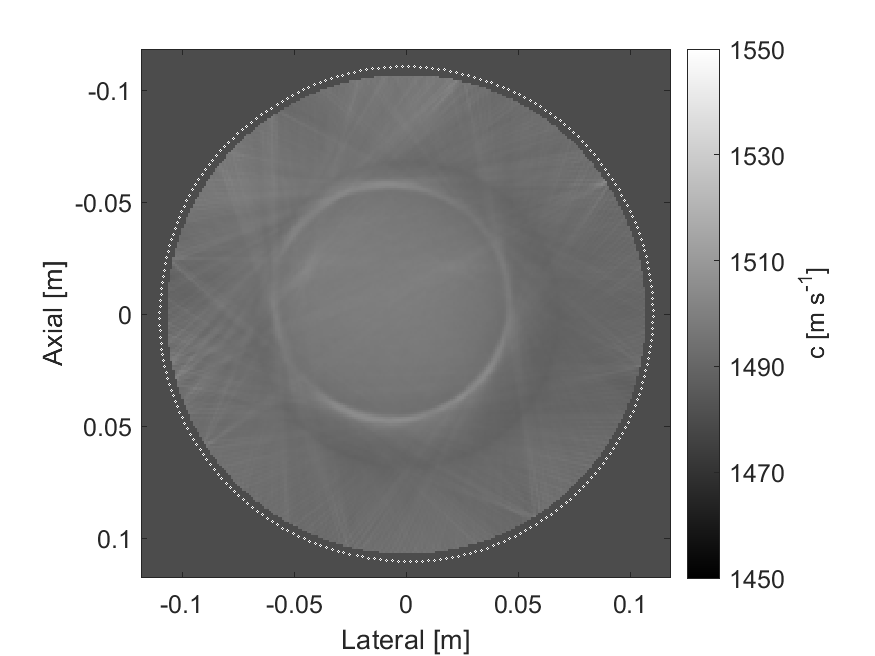}\label{fig:2b}}
    \subfigure[]{\includegraphics[width= 0.33 \textwidth]{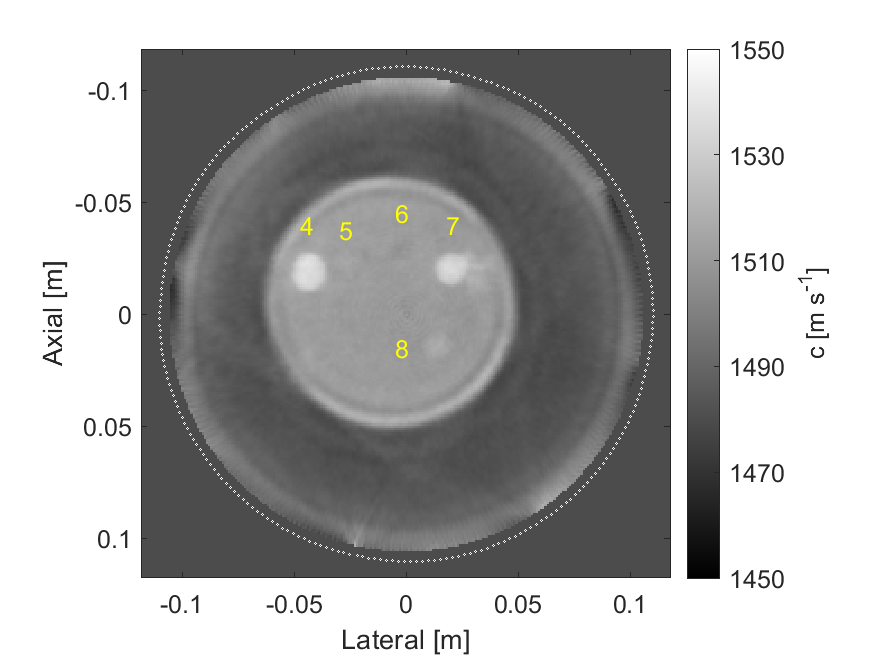} \label{fig:2c}}
    \subfigure[]{\includegraphics[width= 0.33 \textwidth]{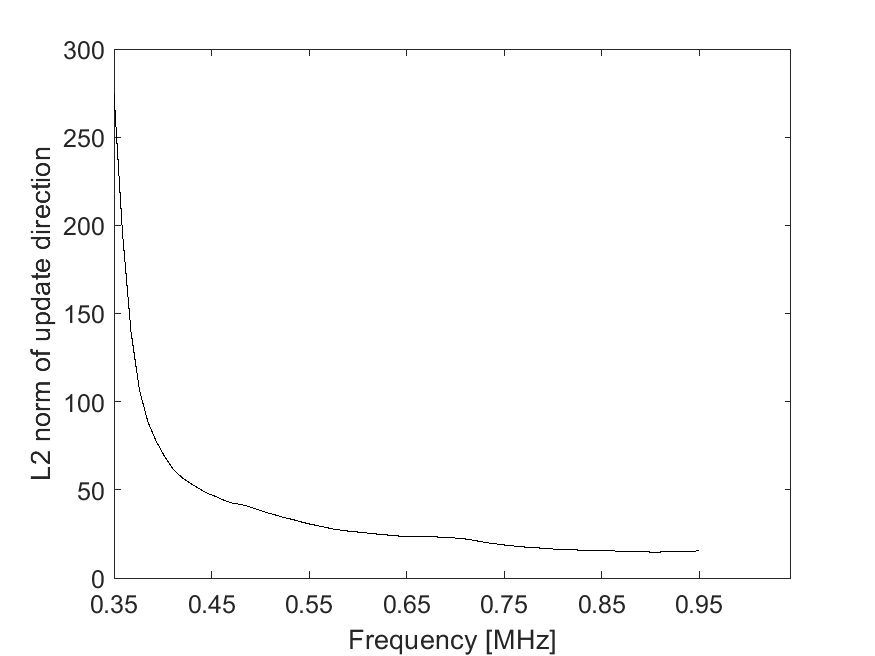} \label{fig:2d}}
   \subfigure[]{\includegraphics[width= 0.33 \textwidth]{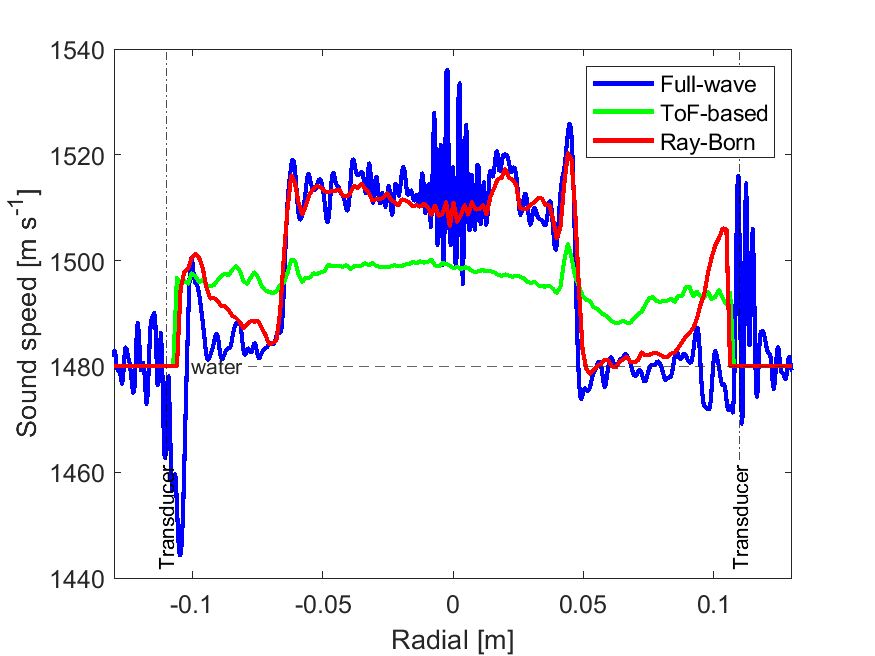} \label{fig:2e}}  
      \subfigure[]{\includegraphics[width= 0.33 \textwidth]{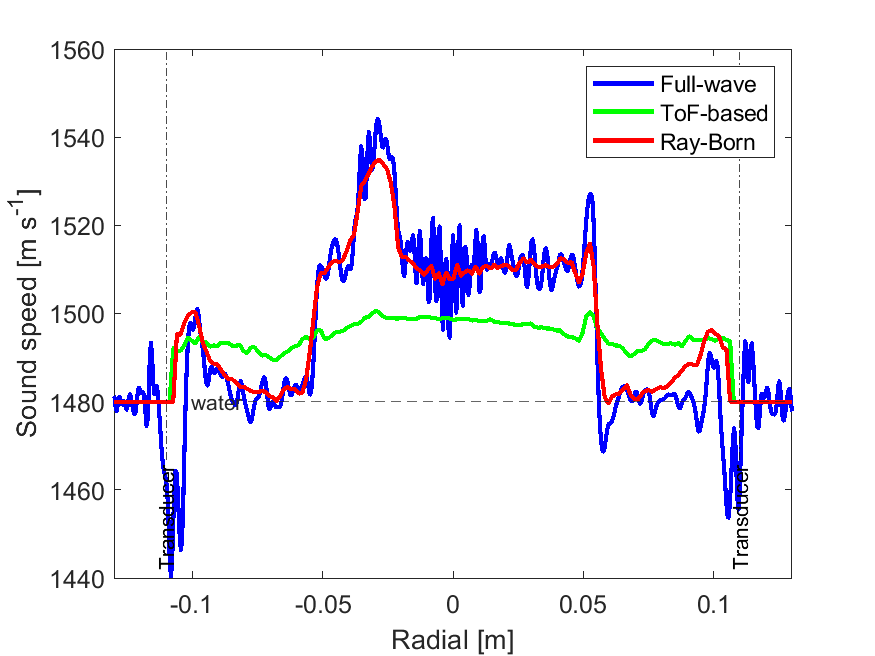} \label{fig:2f}}
       \subfigure[]{\includegraphics[width= 0.33 \textwidth]{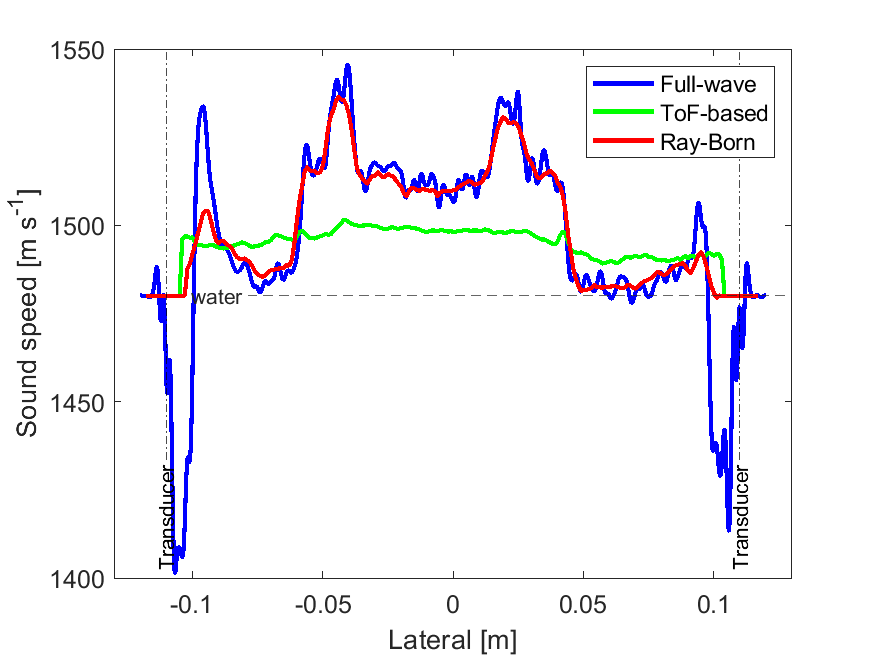} \label{fig:2g}}  
          \subfigure[]{\includegraphics[width= 0.33 \textwidth]{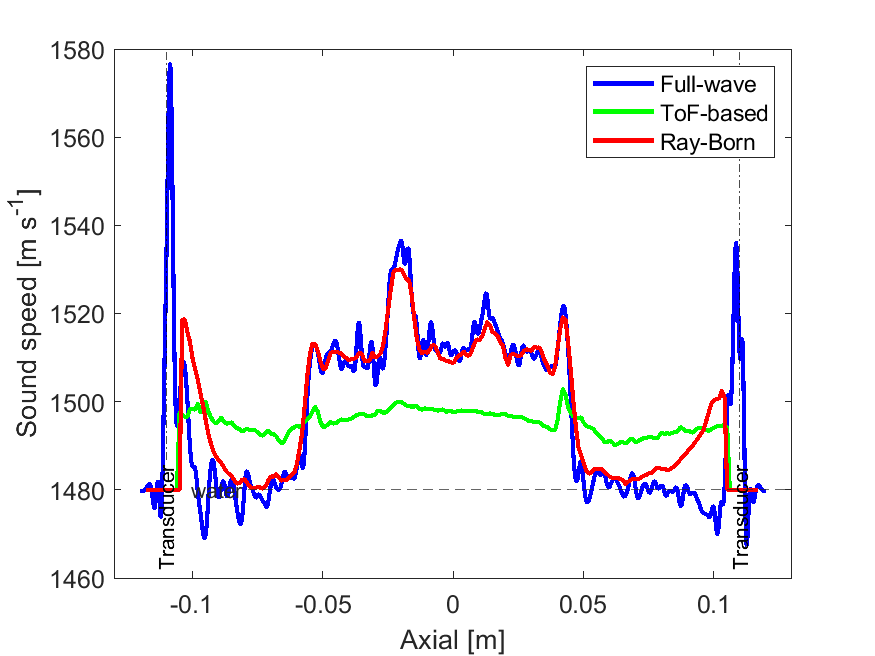} \label{fig:2h}}  
\end{center}
\caption{ \label{fig:2}
Sound speed images from the \textit{in vitro} dataset corresponding to slice~(b) of the phantom, reconstructed using: (a) full-wave inversion, (b) the time-of-flight (ToF)-based approach, which served as the initial guess, and (c) the ray-Born inversion approach. (d) The \(L_2\)-norm of the update direction in terms of sound speed for each frequency subproblem of the implemented ray-Born inversion. Line profiles of the reconstructed sound speed images along: (e) the first diagonal, (f) the second diagonal, (g) the \(x=-2.5 \ \mathrm{cm}\) (lateral) axis, and (h) the \(y=1.5 \ \mathrm{cm}\) (axial) axis.
}

\end{figure}

Similar to Figure~\ref{fig:1}, Figure~\ref{fig:2} presents the quantitative results corresponding to slice~(b) of the phantom described in Section~IV.C of \cite{Ali}. Compared to slice~(a), the low contrast and small size of inclusions~5, 6, and 8 pose a challenge for image reconstruction. A visual comparison of Figures~\ref{fig:2a} and~\ref{fig:2c} shows that the ray-Born inversion approach successfully reconstructs all five inclusions (numbered~4--8), although inclusion~5 appears slightly faded in the reconstructed images for both the full-wave and ray-Born inversion approaches.

Figures~\ref{fig:2e}--\ref{fig:2h} show the line profiles of the reconstructed sound speed images along the first diagonal, second diagonal, \(x=-2.5 \ \mathrm{cm}\) (lateral) axis, and \(y=1.5 \ \mathrm{cm}\) (axial) axis, respectively. The horizontal dotted line indicates the sound speed in water.

Figure~\ref{fig:2d} illustrates the convergence of the ray-Born algorithm, demonstrating that the reconstructed image shown in Figure~\ref{fig:2c} and the line profiles depicted in Figures~\ref{fig:2e}--\ref{fig:2h} are robust with respect to the choice of termination frequency.

\subsection{\textit{In-Vivo} Breast Datasets}

For data acquisition, a total of 1024 transducers were employed, half of which are included in the open-source package \cite{urmc_data}, so the image reconstructions are performed using 512 transducers. The sound speed in water was measured during the experiment, and was 1480~m/s. Ultrasound data were acquired for two cases: (c) a benign cyst; and (d) an example of a spiculated malignancy, as described in Section~IV.D of \cite{Ali}. While the full-waveform inversion approach was performed on a grid with a spacing of 0.3~mm,  the ray-Born image reconstruction approach proposed in \cite{Javaherian_2022} was executed on a coarser grid with a spacing of 1~mm. 

For each emitter, only the receivers located on the opposite 75\% of the ring were included in the image reconstruction to mitigate directivity effects arising from closely spaced emitter--receiver pairs. In addition, following \cite{Ali}, a single-sided Gaussian window was applied to each time trace to suppress events occurring before the first-arrival time. As described in \cite{Ali}, for both \textit{in vivo} datasets, image reconstruction using the full-wave approach was carried out over a frequency range of 0.3--1.25~MHz, progressing from low to high frequencies, with a water-only initial guess. The ray-Born inversion approach uses a low-resolution, low-contrast image—reconstructed from just three iterations of a ToF-based inversion method—as its initial guess.

\subsubsection{Case (c): Benign Cyst}

In the ray-Born reconstruction, each linear equation~\eqref{eq:linear} was formulated and solved at two consecutive frequencies. For the benign-cyst dataset, the same frequency range as that used for the full-wave inversion was employed. The reconstruction comprised 100 linearization steps, corresponding to 200 frequencies within the supported range. For each frequency set, the largest 1\% of values were set to zero, as described in Section~III.E of \cite{Ali}. 

\begin{figure} [ht]
\begin{center}
    \subfigure[]{\includegraphics[width= 0.33 \textwidth]{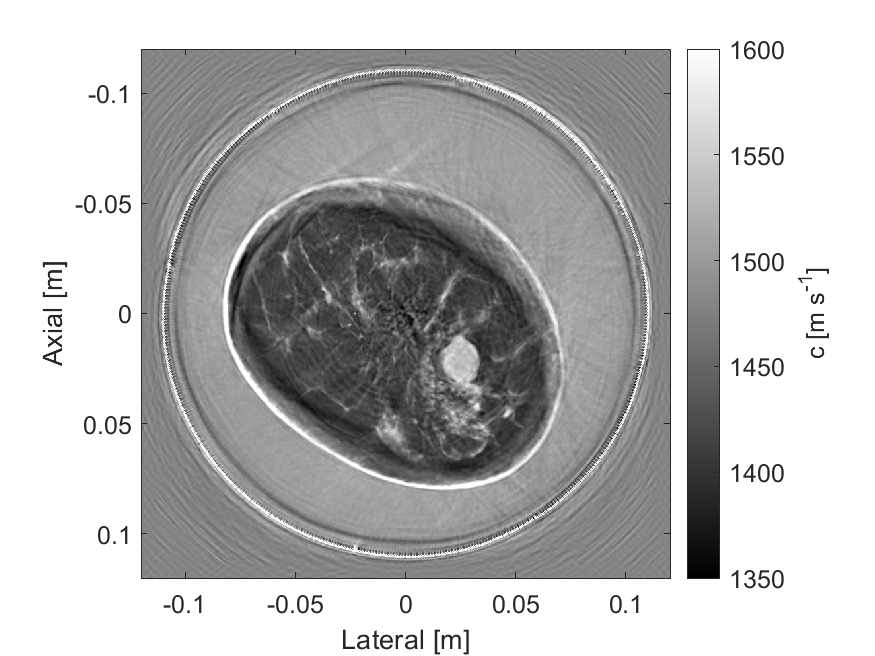} \label{fig:3a}}
    \subfigure[]{\includegraphics[width= 0.33 \textwidth]{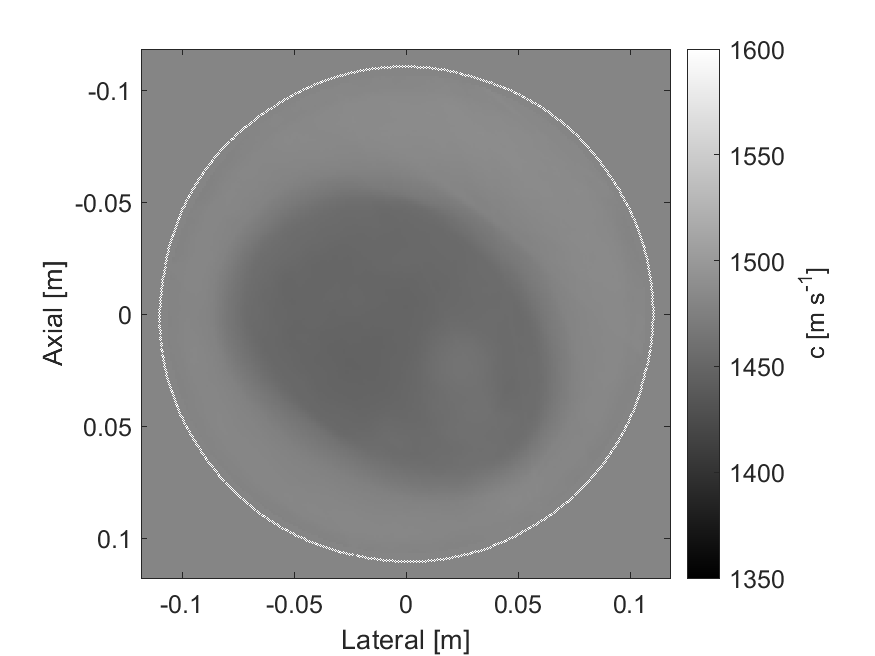}\label{fig:3b}}
    \subfigure[]{\includegraphics[width= 0.33 \textwidth]{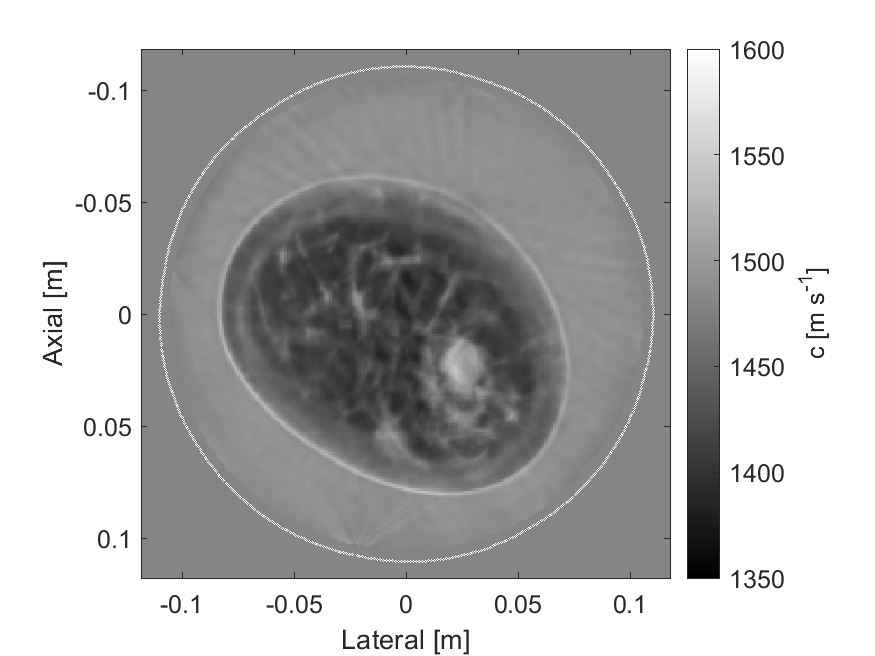} \label{fig:3c}}
    \subfigure[]{\includegraphics[width= 0.33 \textwidth]{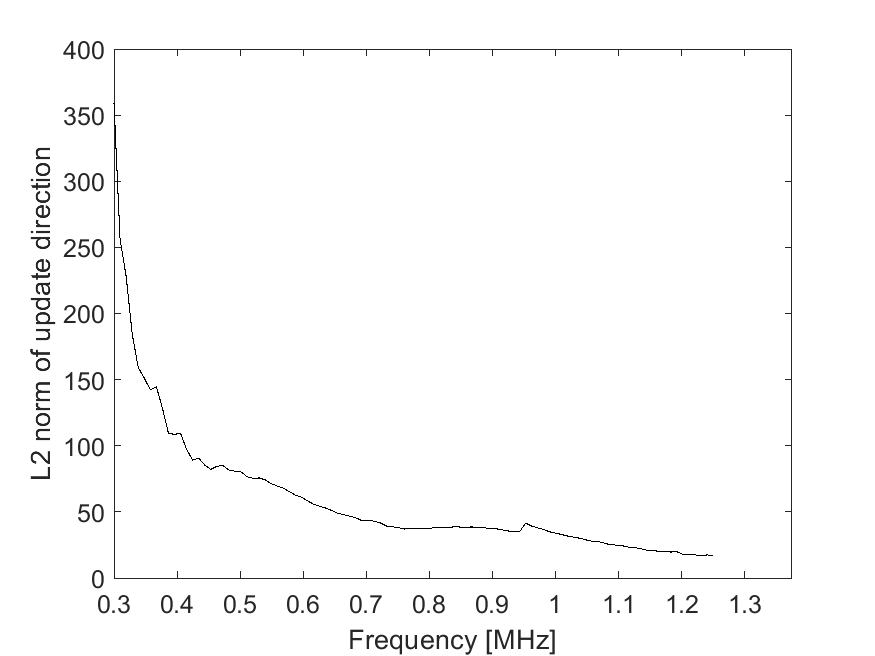} \label{fig:3d}}
   \subfigure[]{\includegraphics[width= 0.33 \textwidth]{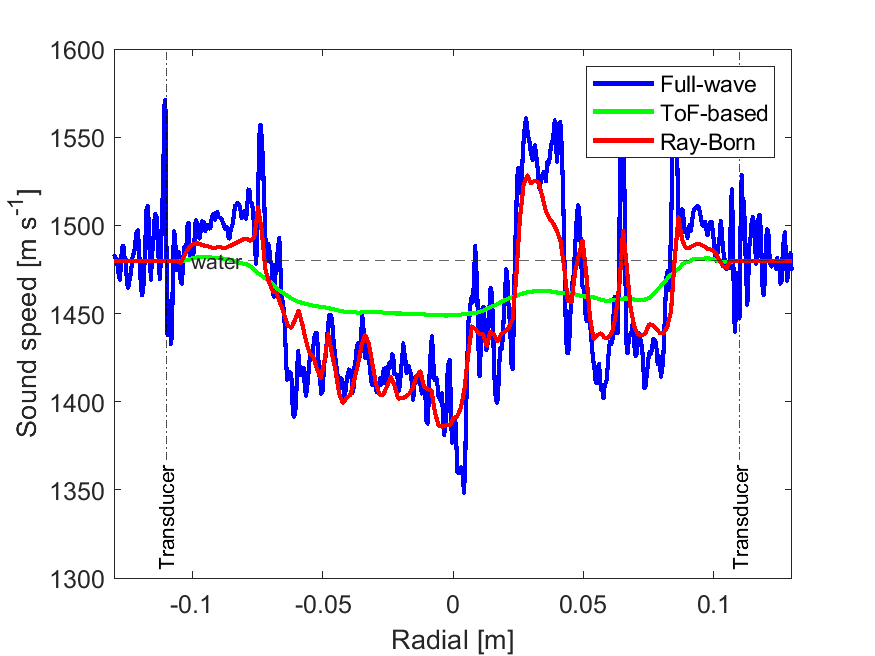} \label{fig:3e}}  
      \subfigure[]{\includegraphics[width= 0.33 \textwidth] {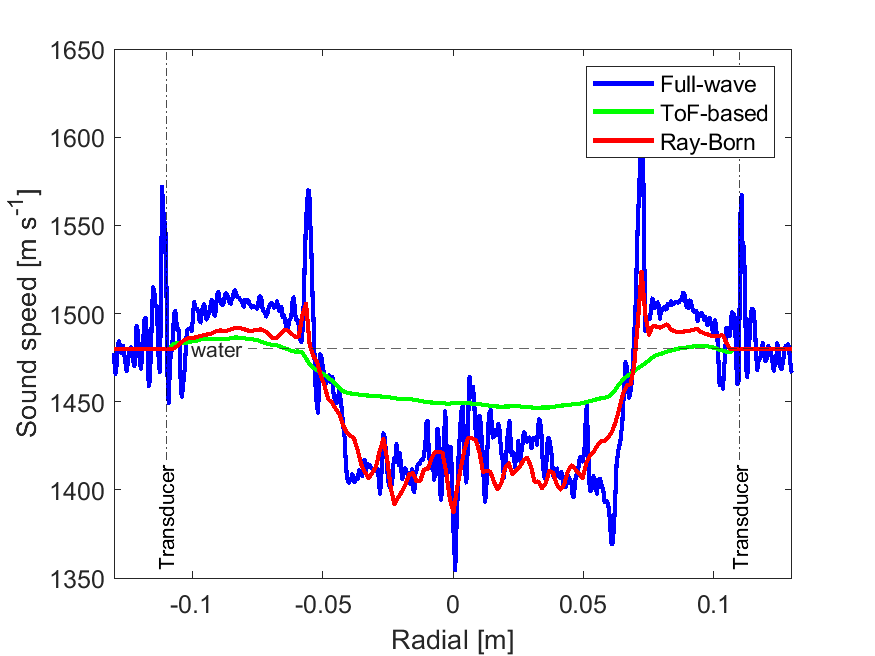} \label{fig:3f}}
       \subfigure[]{\includegraphics[width= 0.33 \textwidth]{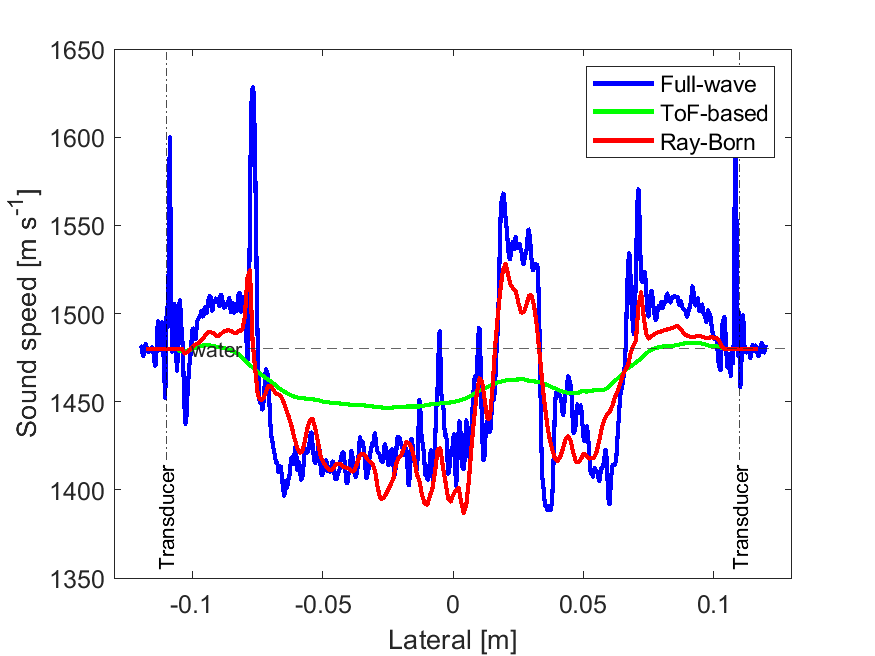} \label{fig:3g}}  
          \subfigure[]{\includegraphics[width= 0.33 \textwidth]{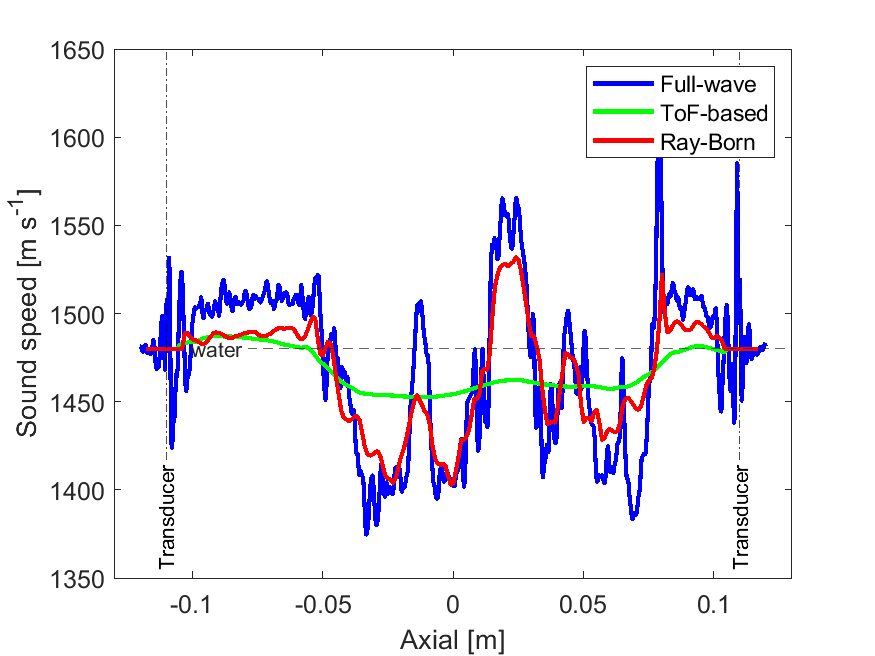} \label{fig:3h}}  
\end{center}
   \caption 
   { \label{fig:3} 
Sound speed images from the \textit{in-vivo} benign-cyst dataset, reconstructed using: (a) full-wave inversion, (b) the time-of-flight (ToF)-based approach, which served as the initial guess, and (c) the ray-Born inversion approach. (d) The \(L_2\)-norm of the update direction in terms of sound speed for each frequency subproblem of the implemented ray-Born inversion. Line profiles of the reconstructed sound speed images along: (e) the first diagonal, (f) the second diagonal, (g) the \(x = 2 \ \mathrm{cm}\) (lateral) axis, and (h) the \(y =  2 \ \mathrm{cm}\) (axial) axis.
}
\end{figure}

Figures~\ref{fig:3a}, \ref{fig:3b}, and~\ref{fig:3c} show the images reconstructed using the full-wave, ToF-based, and ray-Born inversion approaches, respectively. A comparison between Figures~\ref{fig:3a} and~\ref{fig:3c} demonstrates that the ray-Born inversion approach reconstructs small-scale features with lower contrast but fewer artifacts than the full-wave inversion approach. As shown in Figure~\ref{fig:3d}, the \(L_2\)-norm of the update directions decreases progressively from low to high frequencies, demonstrating the robustness of the reconstruction with respect to the termination frequency. 

Figures~\ref{fig:3e}--\ref{fig:3h} show the sound speed profiles along the first diagonal, second diagonal, \(x= 2 \ \mathrm{cm}\) (lateral) axis, and \(y= 2 \ \mathrm{cm}\) (axial) axis, respectively. The horizontal dotted line corresponds to the sound speed in water. From these plots, while the ray-Born inversion approach reconstructs small-scale features with lower contrast, it effectively suppresses artifacts. Furthermore, the ray-Born approach provides a more accurate estimation of the sound speed in water compared to the full-wave inversion approach.

\subsubsection{Case (d): Spiculated Malignancy}

In the ray-Born reconstruction, each linear equation~\eqref{eq:linear} was formulated and solved at two consecutive frequencies. For the spiculated-malignancy dataset, frequencies ranging from 200~kHz to 930~kHz were used in the ray-Born image reconstruction. Compared to the benign-cyst dataset, lower frequencies were used since the single-scattering assumption may progressively lose validity at higher frequencies due to the increased structural complexity and density of breast tissue. Correspondingly, the reconstruction comprised 70 linearization steps, corresponding to 140 frequencies within the supported range. For each frequency set, the largest 1\% of values were set to zero, as described in Section~III.E of \cite{Ali}. 

\begin{figure} [ht]
\begin{center}
    \subfigure[]{\includegraphics[width= 0.33 \textwidth]{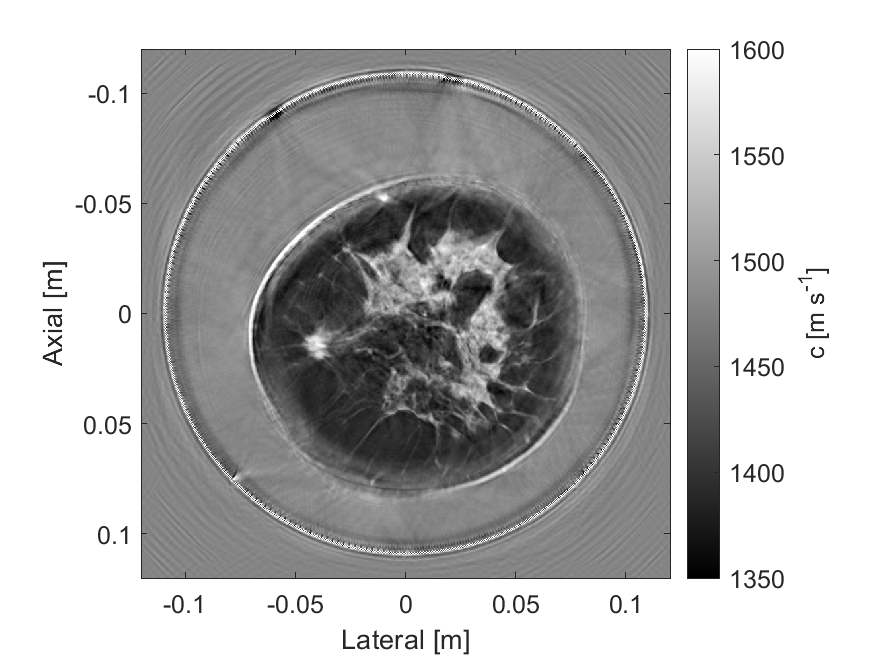} \label{fig:4a}}
    \subfigure[]{\includegraphics[width= 0.33 \textwidth]{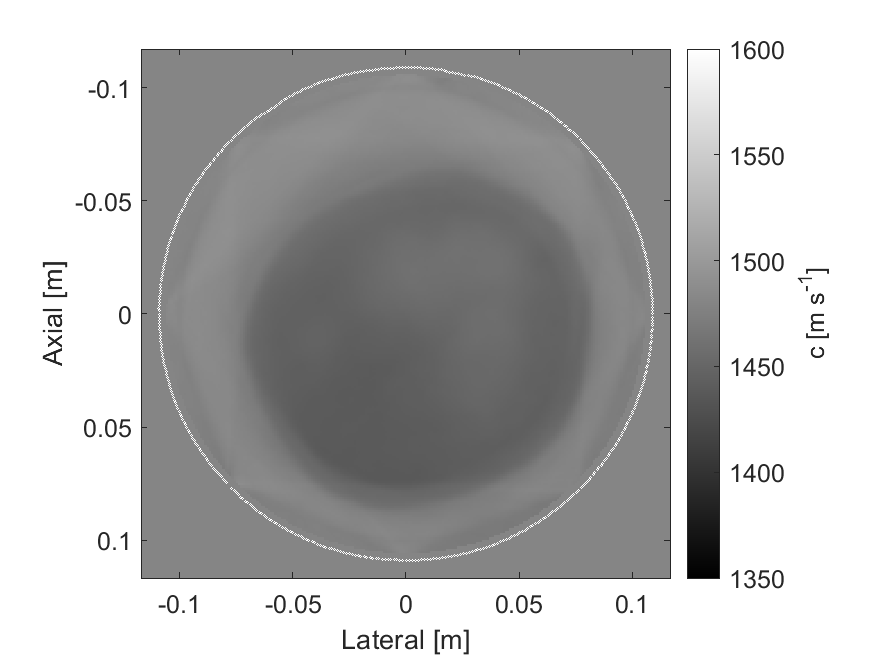}\label{fig:4b}}
    \subfigure[]{\includegraphics[width= 0.33 \textwidth]{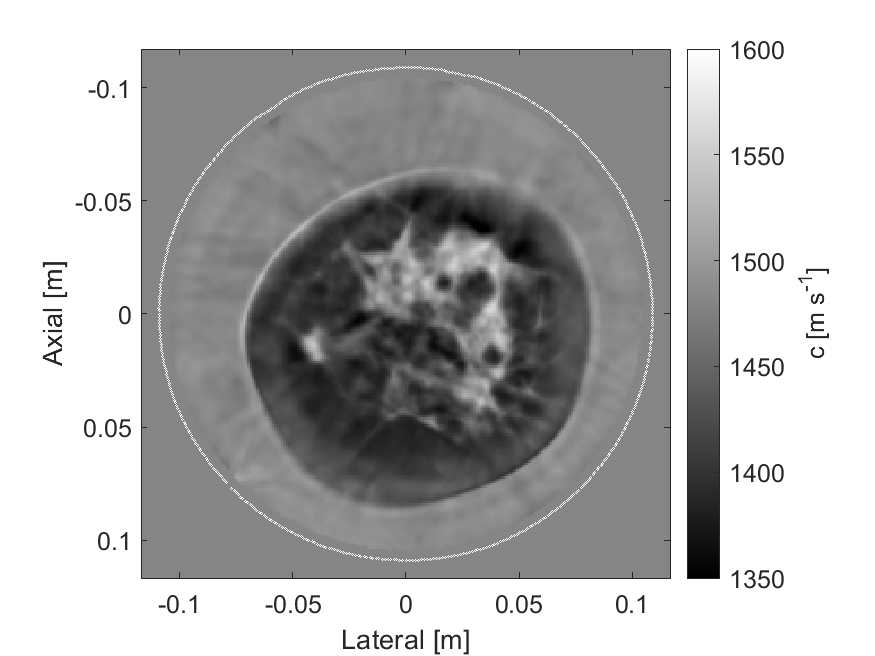} \label{fig:4c}}
    \subfigure[]{\includegraphics[width= 0.33 \textwidth]{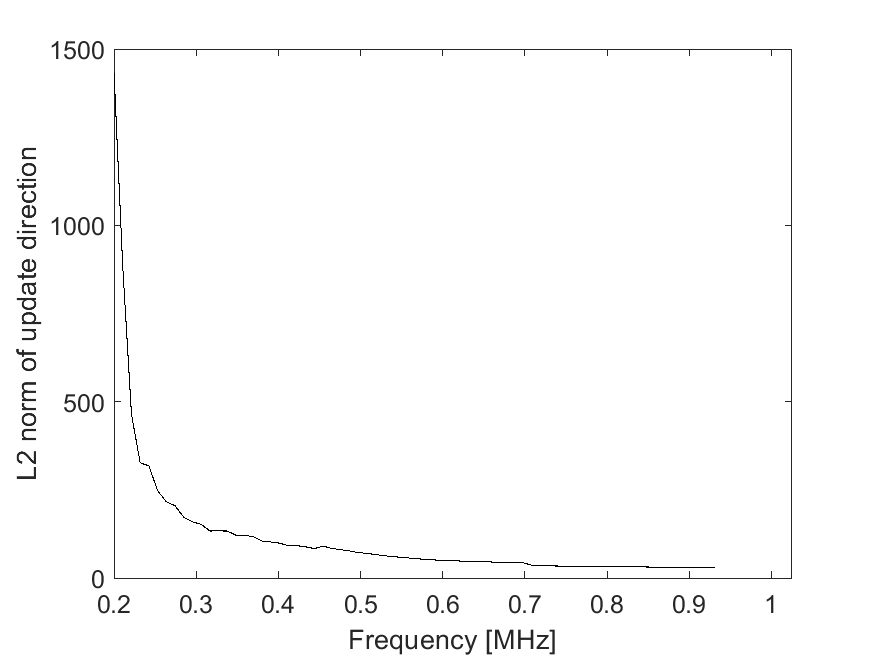} \label{fig:4d}}
   \subfigure[]{\includegraphics[width= 0.33 \textwidth]{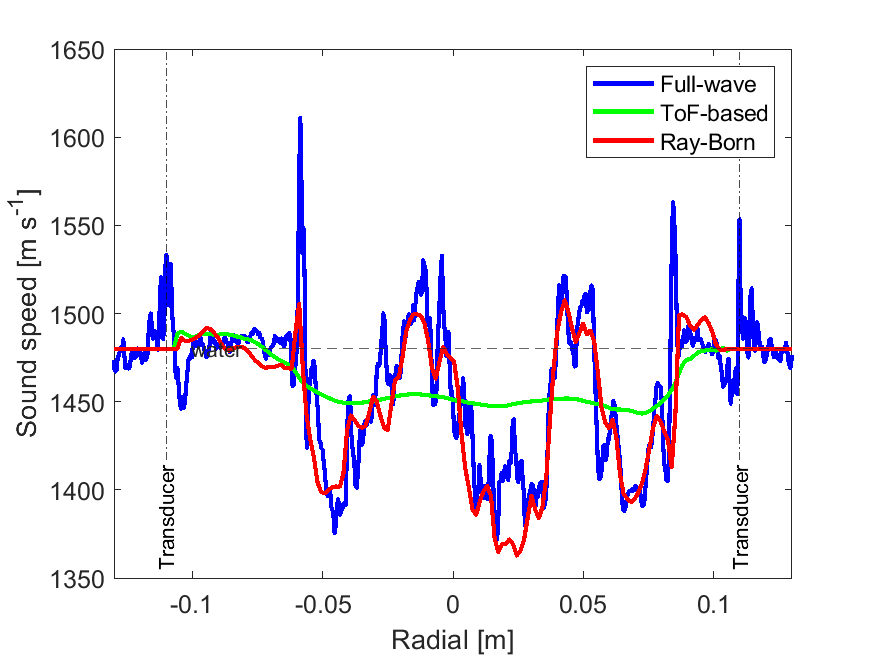} \label{fig:4e}}  
      \subfigure[]{\includegraphics[width= 0.33 \textwidth] {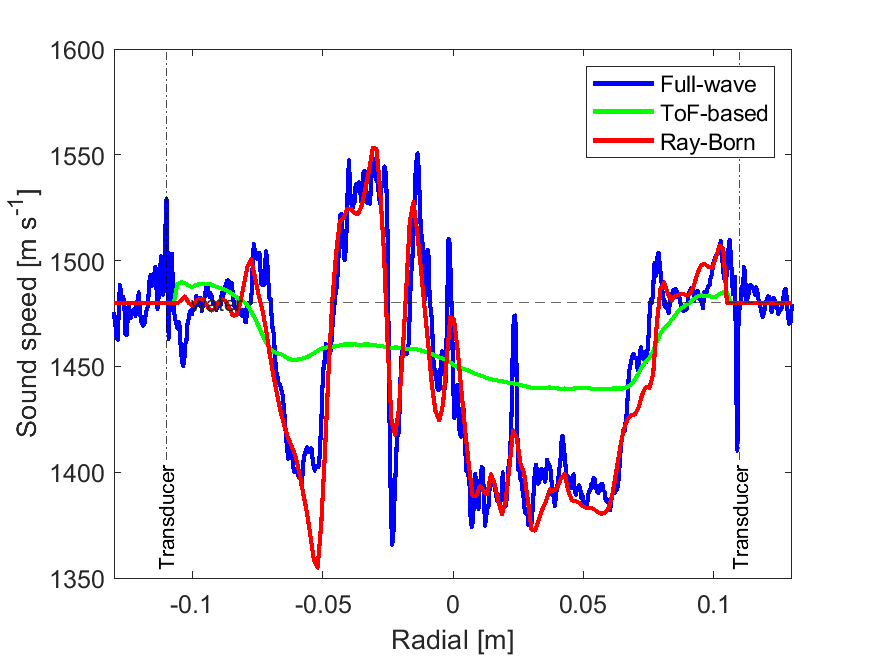} \label{fig:4f}}
       \subfigure[]{\includegraphics[width= 0.33 \textwidth]{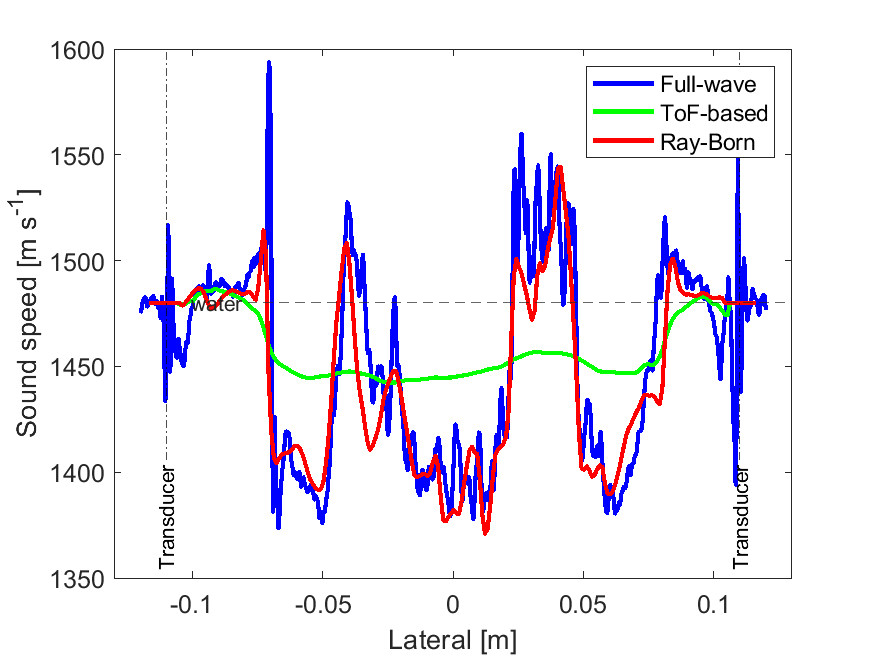} \label{fig:4g}}  
          \subfigure[]{\includegraphics[width= 0.33 \textwidth]{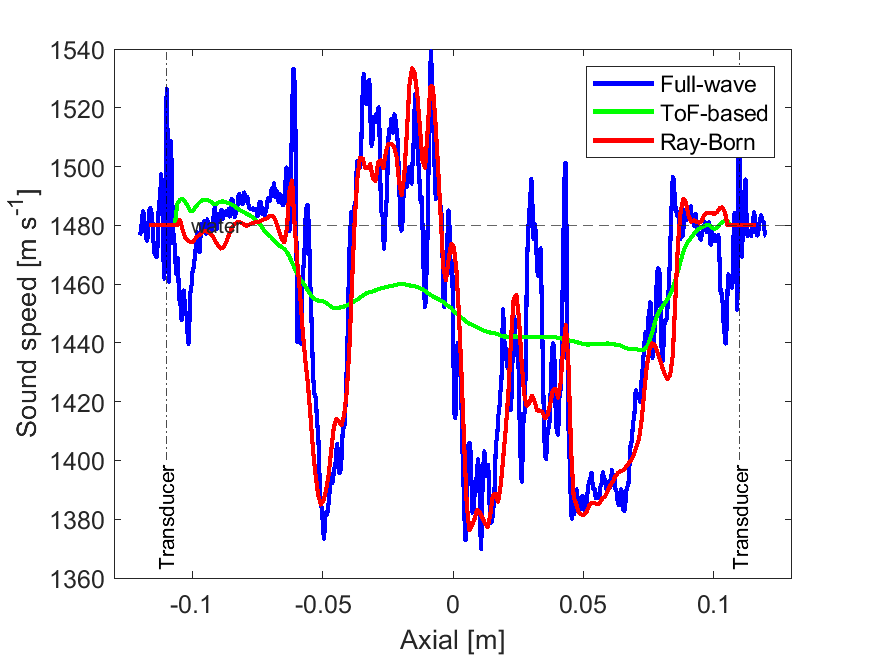} \label{fig:4h}}  
\end{center}
 \caption{
\label{fig:4}
Sound speed images reconstructed from the \textit{in vivo} spiculated-malignancy dataset using: (a) full-wave inversion, (b) the time-of-flight (ToF)-based approach, which served as the initial guess, and (c) the ray-Born inversion approach. (d) The \(L_2\)-norm of the update direction in terms of sound speed for each frequency subproblem of the implemented ray-Born inversion. Line profiles of the reconstructed sound speed images along: (e) the first diagonal, (f) the second diagonal, (g) the \(x =1 \ \mathrm{cm} \) (lateral) axis, and (h) the \(y = 0 \ \mathrm{cm}\) (axial) axis.
}
\end{figure}

Figures~\ref{fig:4a}, \ref{fig:4b}, and~\ref{fig:4c} show the images reconstructed using the full-wave, ToF-based, and ray-Born inversion approaches, respectively. As shown in Figure~\ref{fig:4d}, the \(L_2\)-norm of the update directions decreases progressively from low to high frequencies, demonstrating the robustness of the ray-Born reconstruction with respect to the termination frequency. 

Figures~\ref{fig:4e}--\ref{fig:4h} present the sound speed profiles along the first diagonal, second diagonal, \(x=1 \ \mathrm{cm} \) (lateral) axis, and \(y=0 \ \mathrm{cm} \) (axial) axis, respectively. The horizontal dotted line marks the sound speed in water. These plots demonstrate that the ray-Born inversion approach accurately reconstructs all features while effectively suppressing artifacts.

\section{Discussion and Conclusion}

This study translated the ray-Born inversion approach proposed in \cite{Javaherian_2022} into a practical framework. As part of an open-source \textsc{Matlab} toolbox for reconstructing sound speed images from transmission ultrasound data using two-point ray tracing, available at \cite{Javaherian_toolbox}, this approach was implemented and tested on the \textit{in vitro} and \textit{in vivo} datasets made publicly available by the University of Rochester Medical Center (URMC) \cite{urmc_data}.

The proposed inversion method successfully reconstructed quantitatively accurate and low-artifact sound speed images from both the \textit{in vitro} and \textit{in vivo} datasets, closely matching the results obtained using a full-waveform inversion approach that employs the Helmholtz equation as the forward model \cite{Ali}.
Note that the images reconstructed using the full-waveform inversion approach are not the ground truth, but are used solely as a benchmark for comparison.

For the \textit{in vitro} datasets, the ray-Born inversion approach accurately reconstructed the low-contrast and small inclusions, with their locations, shapes, and quantitative values consistent with those obtained using the full-wave inversion approach.

For the \textit{in vivo} datasets, the ray-Born inversion approach successfully reconstructed fine-scale features whose locations, shapes, and quantitative values were also in agreement with those reconstructed using the full-wave inversion approach.

Overall, the images reconstructed using the ray-Born inversion approach exhibit slightly lower contrast but fewer artifacts than those obtained from full-wave inversion, highlighting its robustness and computational efficiency for practical imaging scenarios.

Image reconstruction approaches that account for singly scattered waves have been extensively studied by engineers and mathematicians for applications in acoustics. However, their translation to clinical settings has remained limited \cite{Simonetti,Huthwaite}. One major limitation is that prototype singly-scattering-based inversion methods typically incorporate the medium’s heterogeneity only in the scattering potential \cite{Guo,Stahli,Martiartu,Kirisits}, while the Green’s functions—responsible for propagating the wavefield between the scattering potentials and the transducer positions—are often approximated under a water-only assumption. This simplification neglects spatial variations in the acoustic properties of the medium, which can lead to poor convergence of the inversion algorithm in practical scenarios or numerical studies that account for realistic physical effects such as absorption and dispersion. Moreover, it can lead to misleadingly stable reconstructions if the so-called \textit{inverse crime} \cite{Wirgin} is inadvertently committed when generating synthetic data used as a benchmark for image reconstruction (see also the \textit{Results} section in \cite{Javaherian_2022}).

Motivated by this limitation, the ray-Born approaches were proposed, combining the Born approximation—valid at low frequencies—with ray theory, which is valid in the high-frequency regime, thereby adapting the method to the frequency ranges used in medical ultrasound tomography \cite{Javaherian_2022,Javaherian_2021}. Unlike methods relying solely on a water assumption, the Green’s functions in the ray-Born approach are approximated along rays whose trajectories are determined using a high-frequency approximation \cite{Javaherian_2022,Javaherian_2021,Cerveny}.

Although the ray-Born inversion is conceptually simpler than full-wave inversion and avoids solving the Helmholtz equation on a dense grid, its current MATLAB implementation does not yet achieve optimal runtime performance. This limitation arises primarily from the fine-grained nature of the computations: the algorithm performs numerous small tasks, such as ray tracing, ray linking, amplitude computation, and back-propagation of ray parameters. In MATLAB, these operations involve frequent data transformations and memory transfers between processing units, which are not efficiently handled by the interpreter. By contrast, full-wave inversion methods rely on large, contiguous matrix operations executed within highly optimized, compiled libraries, making them more cache-friendly and computationally efficient. A compiled implementation of the ray-Born approach in C++ or CUDA, employing batched ray propagation and improved data locality, is expected to overcome these limitations and reveal the intrinsic computational advantage of the proposed method. Despite these limitations, all current reconstructions using the ray-Born inversion approach were performed on the CPU described in the first paragraph of the Results section.

In contrast to full-wave image reconstruction methods, which operate largely as black-box procedures, the ray-Born inversion approach proposed in \cite{Javaherian_2022} comprises multiple modular steps, each of which offers opportunities for refinement that could further enhance image quality—particularly on grids finer than 1 mm. Investigation of these improvements is left for future work.

The present study represents a significant step toward the clinical translation of singly-scattering-based inversion approaches for quantitative reconstruction of sound speed from transmission ultrasound data. By producing low-artifact images, the proposed ray-Born inversion method \cite{Javaherian_2022} offers a promising alternative for hybrid imaging modalities, in which one modality leverages the acoustic heterogeneity determined by another \cite{Javaherian_2019,PATTYN2021100275}. The ray-based approach is particularly advantageous in this context, as reconstructing \textit{a priori} sound speed information via full-waveform methods entails a significantly higher computational cost, while the reduction of artifacts may be more critical than high spatial resolution for such applications.

\section*{Data/code availability}

The MATLAB codes supporting the findings of this study, as well as those in \cite{Javaherian_2022,Javaherian_2021,Javaherian_2020}, are publicly available at the following GitHub repository: \\
\url{https://github.com/Ash1362/ray-based-quantitative-ultrasound-tomography/}.

The in vitro and in vivo datasets used in this work are available at: \\
\url{https://github.com/rehmanali1994/WaveformInversionUST/}.

\section*{Acknowledgments}
I would like to express my sincere gratitude to Professor Nebojsa Duric and his team at the University of Rochester Medical Center, as well as to Delphinus Medical Technologies, for publicly releasing the in vitro and in vivo datasets used in this paper. I also gratefully acknowledge Professor Mohammad Mehrmohammadi, Dr. Rehman Ali, and Mr. Gaofei Jin for their valuable advice and for sharing their experience with experimental ultrasound data.

\bibliography{report} 
\bibliographystyle{spiebib} 

\end{document}